\documentclass[useAMS,usenatbib]{mn2e}
\usepackage{graphicx}
\usepackage{amsmath,amsfonts,amssymb}
\usepackage{wrapfig}          
\usepackage{graphicx}         
\usepackage{epsfig}
\usepackage{psfrag}
\usepackage{pstricks,pst-plot} 
\newcommand{\vek}[1]{\boldsymbol{#1}}
\begin{document}

\title[Ruling out Kozai resonance in PSR~J$1903+0327$ ]
{Ruling out Kozai resonance in highly eccentric galactic binary millisecond pulsar PSR~J$1903+0327$}

\author[Gopakumar, Bagchi \& Ray]
{\parbox[t]{\textwidth}{Achamveedu Gopakumar$^{1,2,*}$, Manjari Bagchi$^{1}$, Alak Ray$^{1}$}\\
\vspace*{3pt} \\
\\ $^1$ Tata Institute of Fundamental Research, Homi Bhabha Road, Colaba, Mumbai 400 005, India.
\\ $^2$ Theoretisch-Physikalisches Institut, Friedrich-Schiller-Universit\"  at Jena, Max-Wien-Platz 1,07743 Jena, Germany 
\\ $^*$ gopu@tifr.res.in}
\maketitle

\begin{abstract}
We investigate the observational signatures associated with
one of the proposed formation scenario for the recently discovered highly eccentric binary millisecond pulsar (MSP) PSR~J$1903+0327$ in the galactic plane. The scenario requires that the MSP to be part of a hierarchical triple (HT), consisting of inner and outer binaries, experiencing the Kozai resonance.  Numerical modeling of a bound point mass HT, while incorporating the effects due to the quadrupolar interactions between the binary orbits and dominant contributions to the general relativistic periastron precession in the inner binary, reveals that, at the present epoch, the orbital eccentricity of the binary MSP should decrease for reasonable ranges in the HT parameters. The estimated decrements in the orbital eccentricity of the inner binary are few parts in $10^{5}$, substantially higher than the reported accuracies  in the estimation of the orbital eccentricity of the binary MSP, while employing various general relativistic timing models for isolated binary pulsars. For wide ranges in the allowed orbital parameters, the estimated rate of change in the eccentricity of the inner binary is orders of magnitude higher than the value recently measured by the pulsar timing analysis. Therefore, we rule out the scenario that the MSP is part of a HT undergoing the Kozai oscillations. The origin of this system in a typical globular cluster is also shown to be less likely than inferred in the discovery paper. 

\end{abstract}

\begin{keywords}
{stars: neutron --- pulsars: individual (PSR J1903+0327) -- methods: analytical -- stellar dynamics -- celestial mechanics}
\end{keywords}

\section{Introduction} 
 
The recent discovery of a MSP PSR~J$1903+0327$ in a highly eccentric orbit around a  solar mass companion in the galactic plane forces one to explore unconventional formation scenarios \citep{djc08}. This is because it is very unusual for a galactic binary pulsar to have spin periods shorter than $10$ ms and a large orbital eccentricity. In the standard model for the formation of the  galactic MSPs, accretion of mass and angular momentum from a binary companion  is required for the spin-up of the pulsar, necessary to create short spin periods \citep{Alpar_82}. Due to the strong tidal effects operating in the binary during the accretion phase,  MSP binaries are expected to be in almost perfect circular orbits \citep{Phinney_92}.

Accurate radio timing of the MSP reveals that the eccentricity of the binary, having an orbital period of around $95$ days, to be $\sim 0.44$ and the reported accuracy of  
its estimation is $\sim 5 $ parts in $10^{9}$, while using a theory-independent general relativistic binary pulsar timing model. Further, the estimated mass of the MSP is $1.74 \pm 0.04 M_{\odot} $.
Infrared observations with Gemini North telescope yielded a possible main-sequence companion star for the MSP. The presence of the main-sequence star in the MSP field prompted its discoverers to suggest triple star interactions to explain the unusually high 
orbital eccentricity of the binary MSP. Let us briefly summarize other plausible formation scenarios for the binary system. Due to its distinct spin and spin-down rates,  Champion $et~al.$ (2008) have almost discarded the  ``born-fast" scenario for this pulsar and strongly favored a binary origin for it. They have also speculated a globular cluster (GC) origin for the system, involving the ejection of the binary MSP from a GC or the disruption of the cluster itself. These cases are also invoked to provide some explanation for the high orbital eccentricity of the system. 

One of the most plausible formation scenario for PSR~J$1903+0327$ requires a HT configuration involving the MSP, its  binary companion, likely to be a  massive white-dwarf, and the main-sequence star, observed in the infrared in the field of the MSP \citep{djc08}. In this scenario, the MSP is part of a primordial HT system composed of the pulsar and the massive white-dwarf. The main-sequence companion star is in a much wider and highly inclined orbit around the inner binary. The  tidal interactions between the inner and outer binaries produce eccentricity oscillations in the inner binary through the Kozai mechanism \citep{K62}. Champion $et~al.$ (2008)  estimate that it is reasonable to expect around two HT systems among the known MSP binaries in the galactic disk, based on an estimate that gave roughly $4\%$ chance for a given neutron star-white dwarf binary to be part of a triple. With the help of Ford, Kozinsky \& Rasio (2000), the authors also noted that it is possible for the inner binary of a highly inclined HT to have eccentricity $\sim 0.44$ during $20\%$ of its Kozai oscillation period, when its masses are $1.74, 1.05$ and $0.9\, M_{\odot}$, respectively. 
These estimates make a HT configuration the most plausible formation scenario for PSR~J$1903+0327$. 

In this article, we explore the physical consequences of Kozai  oscillations operating in the system and its relevance for the on-going timing of  PSR~J$1903+0327$. Using reported values for the orbital  eccentricity and the longitude of periastron, 
extractable from the Table.~I in Champion $et~al.$ (2008),
and reasonable estimates for various parameters necessary to define 
a HT, we demonstrate that, at the present epoch, eccentricity of the binary MSP should decrease. This is achieved by modeling the system to be part of 
a bound point mass HT, while incorporating the effects due to the quadrupolar interactions between the binary orbits and the dominant order general relativistic periastron precession in the inner binary. Typically, the accumulated  eccentricity decrease in an year for the inner binary is substantially higher than the reported accuracy with which the timing eccentricity is estimated (few parts in $10^5$ against few parts in $10^9$). Further, we note that the recently measured  rate of change of the orbital eccentricity $\sim 10^{-16} {\rm s^{-1}} $ (Freire, priv. commun.) unequivocally rule out the above described HT scenario for the MSP binary. We also make an estimate for the spin period derivative, assuming that the MSP is in a triple system and find it to be inconsistent with observations.

Our investigations pertaining to other formation scenarios outlined in Champion $et~al.$ (2008) reveal that the direct observation of the massive white-dwarf companion through its optical band radiation may not be possible in the near future. Additionally,  the origin of this binary MSP in GCs may be subject to more restrictions than has been realized so far. This is because the MSP system like PSR~J$1903+0327$ can originate only from GCs having special properties, thereby reducing the overall probability of this pulsar's origin in the galactic GC systems. 
\section{Essential model to describe the dynamics of Hierarchical triples experiencing the Kozai oscillations} 
\label{HT}

The HT formation scenario for PSR~J$1903+0327$, suggested by its  discoverers, allows us to treat the system to consist of two binaries in quasi-Keplerian orbits that are highly inclined to each other. The inner binary contains the MSP and its stellar-mass companion, possibly a massive white-dwarf, with masses $m_0$ and $m_1$. The stellar-mass main-sequence star having mass $m_2$ forms an outer binary with the center of mass of the inner binary. Let us denote the eccentricities, semi-major axes and the arguments of the periastron (with respect to their lines of nodes) of the inner and outer binaries by $e_1,e_2, a_1, a_2, g_1 $ and $g_2$, respectively and let $i$ be the mutual inclination angle between the two orbits. We probe the temporal evolution of the inner binary with the help of secular perturbation theory, applicable to Newtonian HTs containing point masses, while including the dominant  quadrupolar order  interactions between the two orbits. In other words, the dynamical equations that we invoke are accurate to order $\left ( a_1/a_2 \right )^2$, where $\left ( a_1/a_2 \right )$ is the small parameter in the perturbative expansion. We also incorporate, in an ad-hoc manner, the dominant order general relativistic effect that causes the periastron of an isolated compact binary to advance through the dynamical equation for the argument of periastron of the inner binary. The relevant equations providing secular temporal evolution for the eccentricity and the argument of periastron
of the inner binary \citep{BLS02} read  
\begin{subequations} 
\label{etgt_Eq}
\begin{align}
\label{dgdt}
\frac{d g_1}{dt} &= 
\frac{ 6\,C_2}{G_1} \biggl \{ 4\theta^2+(5\cos2g_1-1)(1-e_1^2-
                 \theta^2) \biggr \} 
\nonumber\\
 &
+\frac { 6\, C_2\, \theta }{  G_2} \biggl \{ 2+e_1^2(3-5\cos2g_1) \biggr \} 
\nonumber\\
 &
+{3\over c^2\, a_1\, (1-e_1^2)}\,
\biggl [ \frac {G\, M_i }{ a_1} \biggr ]^{3/2}
\,,\\
\label{dedt}
\frac{d e_1}{dt} & =\frac{ 30\, C_2}{ G_1} \, {e_1(1-e_1^2)}\, (1-\theta^2)\, \sin 2g_1
\,,
\end{align}
\end{subequations}
where $M_i = (m_0 + m_1)$, the total mass of the inner binary and 
$ \theta = \cos i $. The quantity $C_2$ and the magnitudes of the angular momenta $\vek G_1 $ and $ \vek G_2$  of the inner and outer binaries are given by
\begin{subequations} 
\label{c2g12_Eq}
\begin{align}
\label{c2_def}
C_2 &=  \frac{ G\, M_i\, \eta_i} {16\, a_2} \, \frac{ m_2}{ (1-e_2^2)^{3/2} } \, 
\left({a_1\over a_2} \right)^2
\,,\\
G_1 &= \eta_i\, \biggl \{ G\, M_i^3\, a_1\, (1-e_1^2) \biggr \}^{1/2} 
\,,\\
G_2 &=  m_2 \biggl \{ 
\frac{ G\, M_i^2}{ ( M_i + m_2) } \, a_2\, (1-e_2^2) 
\biggr \}^{1/2} 
\,,
\end{align}
\end{subequations}
where $\eta_i$ is the symmetric mass ratio of the inner binary, given by $\eta_i = m_0\, m_1 /M_i^2$. Note that the Newtonian contributions to Eqs.~(\ref{etgt_Eq}) originate from certain `doubly averaged' Hamiltonian, which is derivable from the usual Hamiltonian for a HT at the quadrupolar interaction order \citep{FKR00}.  The `doubly averaged' Hamiltonian, suitable for describing secular (long-term) temporal evolution of a HT, is independent of the mean anomalies of the inner and outer orbits. This implies that their respective conjugate momenta and hence the semi-major axes, $a_1$ and $a_2$, are constants of motion. Moreover, the justifiable neglect of the radiative losses to the orbital energy and angular momentum of the inner binary due to the emission of gravitational waves implies that there are no reactive contributions to $d a_1/dt $ and $d e_1/dt$.

When one neglects general relativistic contributions and let $G_2 >> G_1$ in Eqs.~(\ref{etgt_Eq}), the resulting equations allow an analytic solution \citep{K62}.
It is also possible to construct an approximate integral of motion in terms of $e_1$ and $\theta$ that allows one to classify the dynamical behavior of a HT with the help of trajectories in the the phase space defined by $e_1$ and  $\cos g_1$. It turns out that if the mutual inclination angle is fairly high and in a certain window, the orbital eccentricities experience periodic oscillations over time-scales that are extremely large compared to the respective orbital periods. The above effect, usually referred to as the Kozai resonance, arises due to the tidal torquing between the two orbits. The Kozai resonance can force initially tiny eccentricity of the inner binary to oscillate through a maximum value, given by ${e_1}^{\rm max} \simeq \left( 1 - \frac{5}{3}\, \cos^2 i_0 \right)^{1/2} $ where $i_0$ is the initial value for the mutual inclination angle.
Due to the obvious restriction, namely  $| \cos i_{0} <  (3/5)^{1/2} |$, $i_0$ is required 
to lie in the range $39^{\circ} - 141^{\circ}$ \citep{BLS02}.

The general relativistic periastron advance of the inner binary, in principle, can interfere with the Kozai resonance and even terminate the eccentricity oscillations. 
This is because the extra contributions to $d g_1/dt$ can indirectly affect the evolution of $e_1$. The following useful criterion, derived in Blaes, Lee \& Socrates 2002,
may be used to infer the possibility for this not to happen. 
\begin{equation}
\label{alpha_GR}
 \left ( \frac{ a_2}{a_1} \right ) < \biggl [ \frac{3}{4}\, \frac{ m_2 }{M_i} \, \frac {\tilde a_1} { \tilde M_i } \,
\left ( \frac{ 1 - e_1^2 } { 1 -e_2^2}  \right )^{3/2}  \biggr ]^{1/3} 
\,,
\end{equation}
\newline
where $ \tilde M_i $ is $M_i/M_{\odot}, \tilde a_1 = a_1/L_{\odot} $ and $ L_{\odot} = 1.476625$ km. The above inequality is obtained by equating the right hand side of Eq.~(\ref{dgdt}) to zero, after neglecting the much smaller  $C_2/G_2$ contributions to $d g_1/dt $ and demanding that the resulting expression for $ \cos^2 i$ remains positive. 

Another constraint for $ a_2 / a_1 $ can be obtained by invoking an empherical relation obtained by \citep{MA_01}, relevant while discussing the stability of Newtonian coplanar prograde orbits in HT configurations. The empherical criterion of \citep{MA_01} reads
\begin{equation}
\label{alpha_MA}
 \left ( \frac{ a_2}{a_1} \right ) >  \frac{ 2.8} { 1 -e_2} \, \biggl [  \biggl ( 1 + \frac{ m_2}{M_i} \biggr )\, 
\frac{ ( 1 + e_2) }{ ( 1-e_2)^{1/2} } \biggr ] ^{2/5}
\,.
\end{equation}
Following Blaes, Lee \& Socrates (2002), we treat the above inequality to be rather conservative as the inclined orbits, relevant for our investigation, are expected to more stable than the coplanar triples of Eq.~(\ref{alpha_MA}).

An initial estimate for $e_2$, appearing in the above inequalities, can be obtained by equating the general relativistic periastron precession timescale of the inner binary to the characteristic time scale for the Kozai oscillations. The period for general relativistic periastron precession is deductible from the reported period for advance of periastron $\sim 10^6 $ years, which assumes that the MSP binary is isolated.  An approximate expression for the period of the Kozai oscillation reads
\begin{equation}
\label{t_Kozai}
 \tau_{\rm Kozai}  = P_i \, \frac{ M_i}{m_2}\, \left ( \frac{a_2}{a_1} \right )^3\, ( 1 - e_2^2 )^{3/2} \,,
\end{equation}
where $P_i$ is the orbital period of the inner binary \cite{MS79}. 
With these inputs, for a HT system with $M_i \sim  2.79\,M_{\odot}, a_1 \sim 0.211\, {\rm A.U}, a_2 \sim 600 \times a_1, m_2 \sim 0.9 M_{\odot} $ and $ P_i \sim 0.261$yr, the likely values for a HT listed in Champion $et~al.$ (2008),  the above prescription gives $e_2 \sim 0.979 $. Not surprisingly, the above value for $e_2$ fails to satisfy both the inequalities. However, we observe that a slightly higher value for $e_2$, say $e_2 \sim 0.984$, makes sure that the above two inequalities are satisfied. This is how we prescribe a suitable value for $e_2$ in our computations.

\section{Discussions} 
\label{DISCUSSIONS}

\begin{figure}
\centerline{\psfig{figure=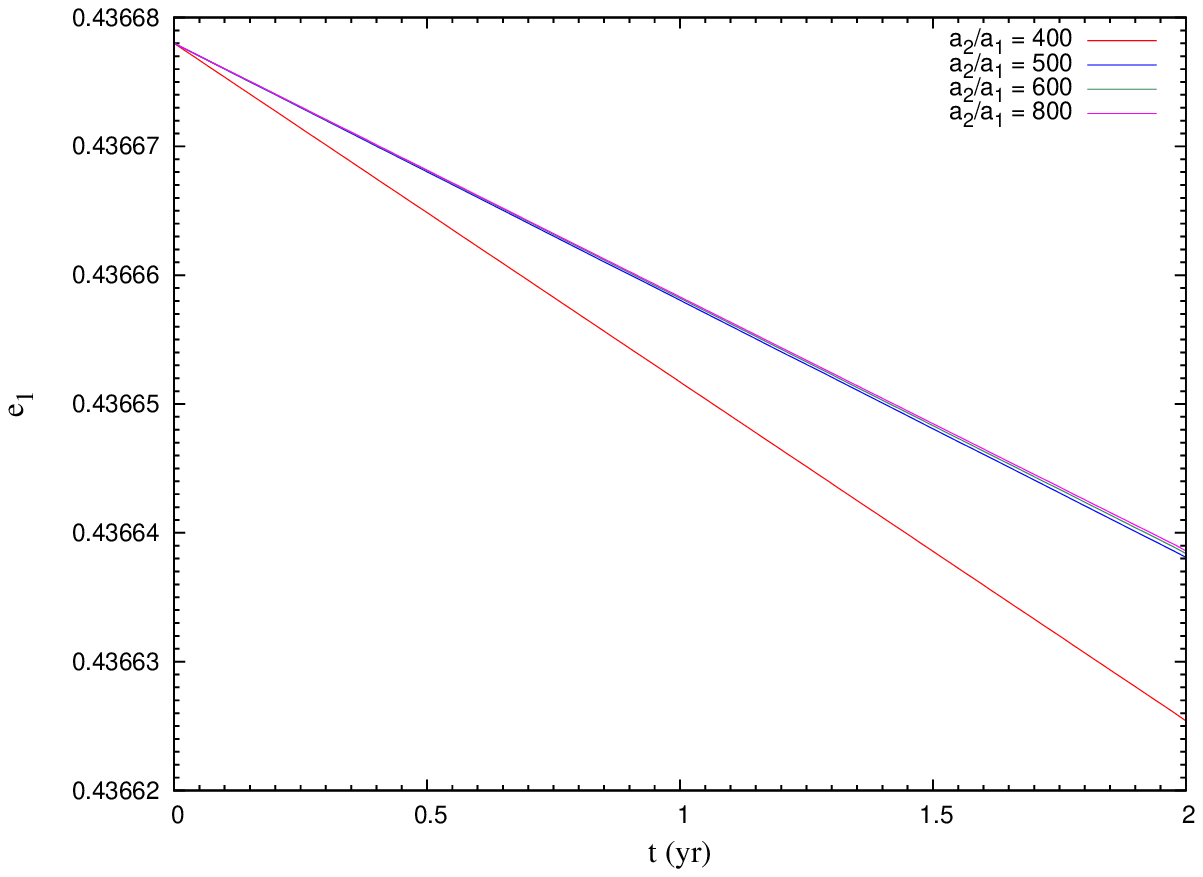,width=8cm}}
\centerline{\psfig{figure=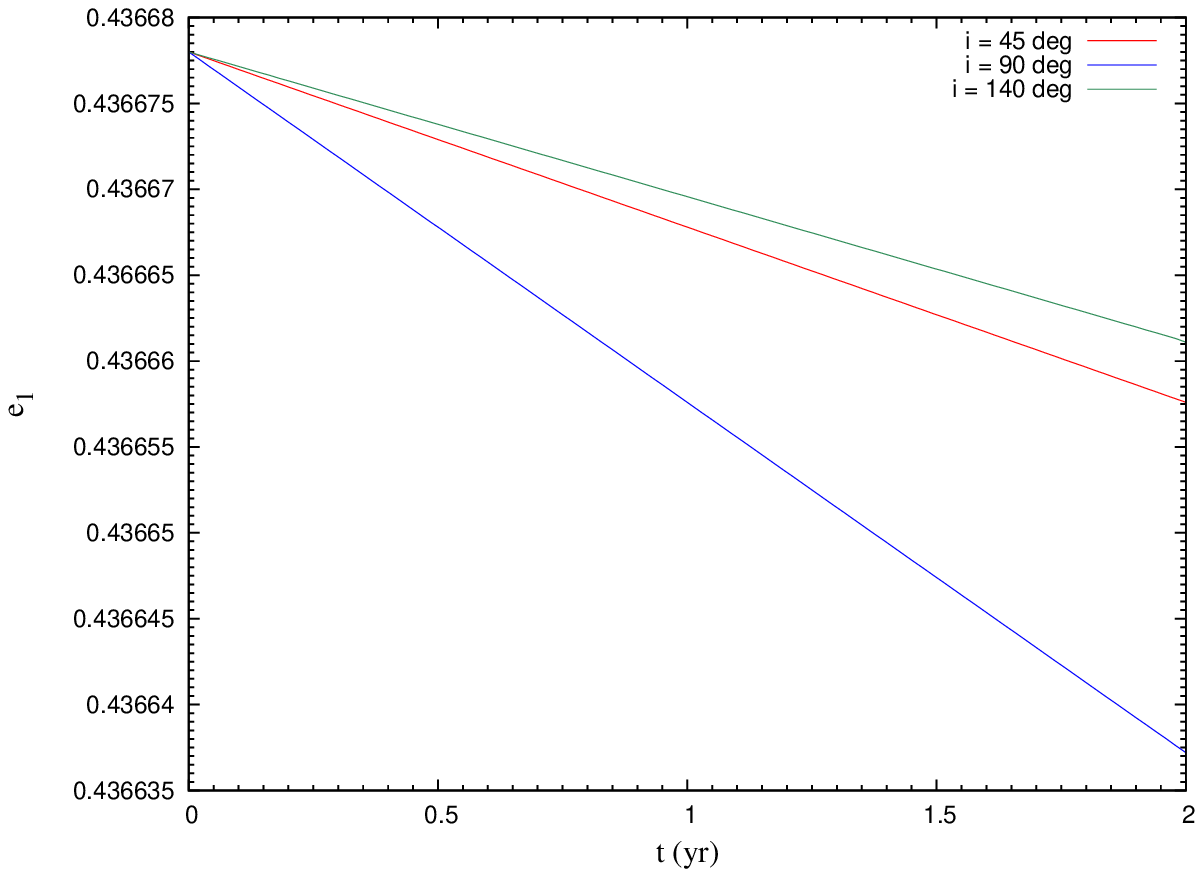,width=8cm}}
\caption{Temporal evolution of the eccentricity of the inner binary over two years for system parameters listed in the text.  In the upper panel, we let $a_2/a_1$ have values 400, 500, 600 and 800 and the associated values for $e_2$ are $ 0.970,0.977, 0.984 $ and $0.991$, respectively. The $e_2$ values are chosen so as to satisfy the two
inequalities, though for the  $a_2/a_1= 800$ case, the conservative inequality for $a_2/a_1$, given by Eq.~(\ref{alpha_MA}), is not satisfied. Note that the  decrements in $e_1$ are fairly independent of $a_2/a_1$ for  $a_2/a_1 >500$
and this may be attributed to the $1/(1 -e_2^2)$ dependency of the right hand side of Eq.~(\ref{dedt}). In the lower panel, the mutual inclination angle between the two binary orbits $i$ takes values $45^{\circ},90^{\circ}$ and $140^{\circ}$ and we fix $a_2/a_1 = 600$ which leads to $e_2$ being $0.984$.  The higher decrement for the $i =90^{\circ}$ case is expected due to more efficient occurring of the Kozai oscillations, 
for which $i$ should be in the range $\sim 39^{\circ}$-$141^{\circ}$. The system parameters in this panel are the same as those for the upper panel.} \label{fig:e_t}
\end{figure}

\begin{figure}
\centerline{\psfig{figure=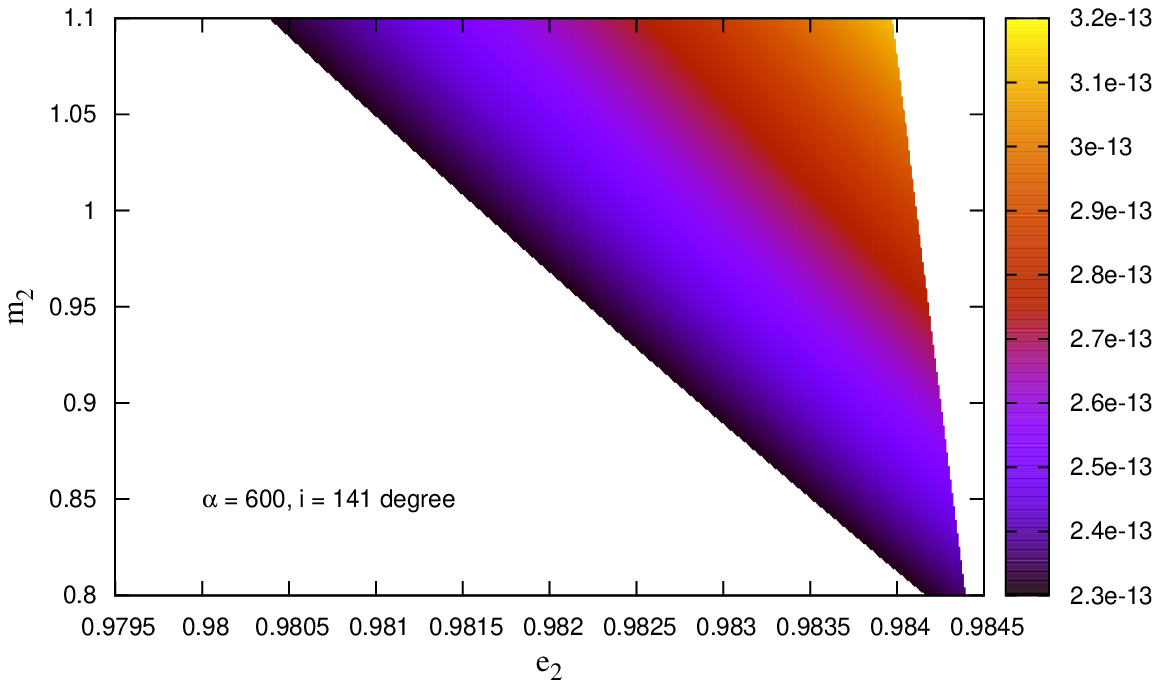,width=8cm}}
\centerline{\psfig{figure=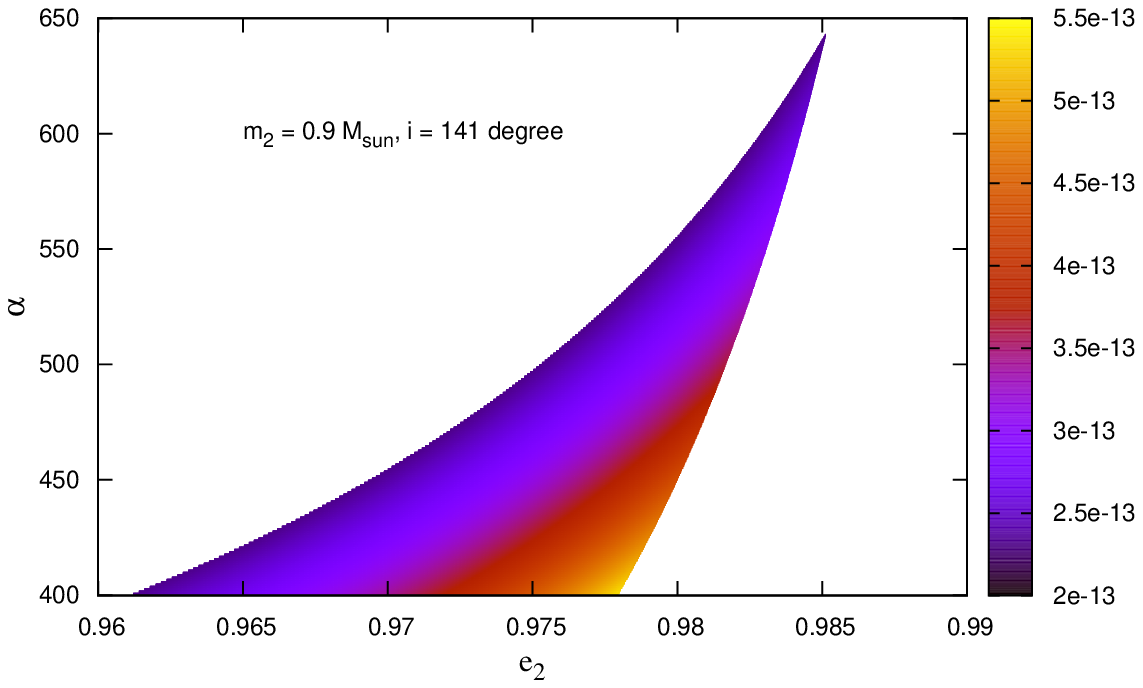,width=8cm}}
\caption{Colour coded contours of $\dot e_1 = d e_1/dt $ in $/{\rm sec}$
arising from Eqs.~(\ref{etgt_Eq}). In the upper panel, we vary $m_2$ and $e_2$
for constant $\alpha$ and $i$ values, while in the lower panel $\alpha$ and $e_2$ are varied with fixed $m_2$ and $i$ values. The coloured regions are defined by the inequalities (\ref{alpha_GR}) and (\ref{alpha_MA}),
providing the left and the right boundaries of the allowed regions, respectively.
The chosen value of $i$ ensures minimum possible $ \dot e_1 $, which is much larger than the recently measured $\dot e$ value for the MSP binary.
 } \label{fig:g_t}
\end{figure}

Let us begin by plotting the temporal evolution in $e_1$, governed by Eqn.~(\ref{etgt_Eq}), that last for couple of years. For Fig.~\ref{fig:e_t}, we adopted the likely system parameters, arising from the radio and infrared observations\citep{djc08}. Therefore, we let $m_0=1.74~M_{\odot}, m_1=1.051~M_{\odot},m_2=0.9~M_{\odot},e_1 =0.436678, a_1 \sim 0.211$ AU and $g_1 = 141.65^{\circ}$.  
Notice that we have employed the values listed in Table.~1 of  Champion $et~al.$ (2008), arising from the `DDGR' timing model \citep{TW}. The other quantities required to obtain $e_1(t)$ are $i, a_2 $ and $e_2$ and as they are not constrained from any observations, we treat them mostly as free parameters in these plots. For a given value of $a_2/a_1$, we estimate $e_2$ such that the inequalities given by Eqs.~(\ref{alpha_GR}) and (\ref{alpha_MA}) are satisfied and $i$ is taken to be in the range so that the inner binary can experience Kozai resonance. The upper panel of Fig.~\ref{fig:e_t} provides $e_1(t)$ plots for four different values of $ a_2 / a_1 $, while keeping $i= 100^{o}$, and in the lower panel, we have  $e_1(t)$ plots for three canonical values of $i$, while keeping $ a_2 / a_1 = 600$. The striking feature of these plots is the decrease in the eccentricity of the inner binary with time and  the decremental change in $e_1$ during a year appears at the \emph {fifth decimal place}. 
From Eq.~(\ref{dedt}), it should be noted that the magnitude of $de_1/dt$, while employing the above mentioned system parameters, depends only on $ \left ( \alpha\, \sqrt { 1-e_2^2} \right )^{-3 }\, ( 1 - \theta^2) $, where $\alpha = a_2/a_1$ and $( 1 - \theta^2)$ is allowed to take values between $1$ and $0.4$ for HT experiencing the Kozai resonance. However, the restriction that any chosen value for $e_2$ should satify the  inequalities,~(\ref{alpha_GR}) and (\ref{alpha_MA}), implies that 
$ \left ( \alpha\, \sqrt { 1-e_2^2} \right )^{-3} $ remains almost a constant ($ \sim 10^{-7}$) for $ a_2/a_1 \ge 500 $. This is reflected in the apparent convergence of $e_1(t)$ curves displayed in  Fig.~\ref{fig:e_t}.  These arguments allow us to state that it is impossible for $de_1/dt$, as deducted from Eqs.~(\ref{etgt_Eq}), to be substantially smaller than $ \sim 10^{-13}{\rm s^{-1}}$ by varying  $a_2/a_1, e_2, i $ and $m_2$ and we depict these conclusions pictorially in Fig.~\ref{fig:g_t}. Further, we note that that whether $e_1$ decreases or increases depends crucially on the initial value for $g_1$ and as $g_1 = 141.65^{\circ}$ at the present epoch, $e_1$ must decrease with time.
  
We observe that the above estimates for the change in $e_1$ over a two year period is substantially higher than the reported precision of the orbital eccentricity for the MSP \citep{djc08}. Moreover, the latest timing results for the PSR~J$1903+0327$ (Freire, priv. commun.) imply that $\dot e \sim 10^{-16}{\rm s^{-1}}$ which is very much smaller than our predictions based on HT configurations for the MSP. Therefore, we rule out the possibility that the MSP binary is part of a HT configuration experiencing Kozai oscillations.

Other measurements can be also used to constrain the orbital configuration for the MSP binary. A rough estimate for the maximum amplitude of the spin period derivative $\dot P$ is obtained by treating the inner binary as a single object of mass $m_0+m_1$ which forms a wide binary with stellar mass companion of mass $m_2$ \cite{JR97}. Further, we let the projected semi-major axis of this new binary to be $600$ times that of the inner binary and choose a value for $e_2$ consistent with inequalities (\ref{alpha_GR}) and (\ref{alpha_MA}). The associated line of sight acceleration leads to an estimate for the maximum amplitude of  $\dot P $ of the order of $ 10^{-17} {\rm s / s}$ and this is much higher than the reported value $\dot P \sim  10^{-20} {\rm s / s }$ of Champion $et~al.$ (2008). This too suggests that the MSP binary is not part of triple system. With the knowledge that the MSP binary is not part of a HT, we note that the evolution of the argument of periastron of the binary can arise only from the third term in Eq.~(\ref{dgdt}). Therefore, the reported rate for the advance of periastron can safely be explained with the help of general relativity. However, the measured periastron advance rate can originate from the classical spin-orbit coupling if the binary companion is a main-sequence star having unusually low rotational period $<0.5$ day \cite{djc08, wex98}.

Champion et.al (2008)  had also discussed the possibility of further optical observations of the main-sequence star to probe the triple nature of the system. Here we briefly explore the feasibility of observing the optical band radiation of the massive white-dwarf companion in the optical image which contains the main-sequence star. Using white dwarf models by L\"ohmer $et~al.$ (2004), we find that a white dwarf of mass $\sim 1 ~M_{\odot}$ and age $ \sim 1.8 \times 10^{9}$ years will have an effective temperature $T_{eff} \sim 10^{4}$ K and an absolute visual magnitude $M_v \sim 13.0$. Taking the extinction coefficient in the visual band ($A_v$) to be 4.9 and the distance $\sim 6.4 $ kpc, employed by Champion et.al (2008), we get an apparent visual magnitude $m_v \sim 31.93$. Using the white-dwarf colors $m_b - m_v$ and $m_b - m_v$, as given in Bergeron $et~al.$ (1995),  and ratios of the extinction coefficients ($ A_b/A_v$  and $A_r/A_v $) from Cardelli $et~al.$ (1989), we get an apparent blue magnitude  $m_b \sim 33.8$ and an apparent red magnitude  $m_r \sim 31$. These high values of apparent magnitudes imply that it will be quite difficult to observe, in the optical band, the companion white dwarf in the near future, even though it has a different spectral characteristic compared to the mainly infra-red optical emission from the main sequence star.

\begin{figure}
\centerline{\psfig{figure=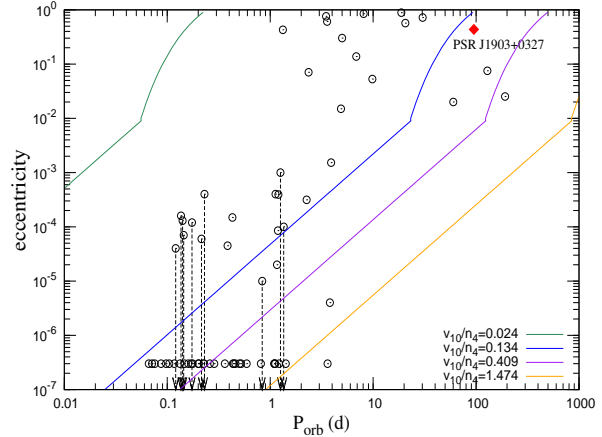,width=8cm, height=6cm}}
\caption{Fly-by isochrones for $t_{fly}=10^{9}$ years in the eccentricity - orbital period plane for different values of the GC parameter $v_{10}/n_4$, where $v_{10}$ is the velocity dispersion of stars in a GC and $n_4$ being the number (central) density of stars.The GC binary pulsars with spin periods $< 20 \;\rm ms$ have been plotted with black $\circ$ symbols and the eccentric galactic disk millisecond pulsar PSR J1903+0327 has been marked with a red $\diamond$.
The region to the lower right of each of these lines have shorter interaction timescales.
This indicates that our binary pulsar could have originated only in a rather special GC environment with characteristics 
belonging to the upper left portion of the isochrone with $v_{10}/n_{4}=0.134$. } \label{fig:gc}
\end{figure}

Let us comment on the possible GC origin for the system, motivated by the fact that more than $15 \%$ of the MSP binaries in GCs are in eccentric orbits and some of them have stellar mass companions \citep{djc08}. We adapt for PSR J1903+0327 the discussions pursued in Bagchi and Ray (2009a, 2009b) to theoretically explain the distribution of orbital eccentricity of binary radio pulsars in GCs in the light of interaction models. In Fig.~\ref{fig:gc}, we plot theoretical separations in the eccentricity-orbital period phase space for binaries that could have acquired a given eccentricity for its corresponding orbital period through fly-by interactions in the past 1 Gyr of its existence. Note that these separations based on lifetimes depend crucially upon the parameter $v_{10}/n_4$ that varies widely between the galactic globular cluster systems by over three orders of magnitude (here $v_{10}$ is the velocity dispersion of stars in a GC while $n_4$ is its central (number) density of stars). The pulsar PSR J1903+0327 like systems could have been produced in GC systems upto a mean  $v_{10}/n_4 = 0.13$ (e.g. a system like NGC 6752 or even better in Ter 5) but not in any globular cluster with a higher value of this parameter. Thus the putative origin of PSR J1903+0327 inside a GC could have taken place only in a restricted set of GCs. This further restricts the environment for production of such pulsars and therefore the overall probability for origin of this pulsar in a GC must be smaller than that estimated by Champion et al (2008). Note that 
the fly-by interactions as well as the exchange and merger interactions scale with the parameter  $v_{10}/n_4$.

\section{Conclusions}
\label{CONCLUSIONS}

We probe the observational implications if  
the recently discovered highly eccentric galactic
binary MSP PSR~J$1903+0327$ is part of a HT experiencing the Kozai oscillations. We model the binary MSP and the main sequence star, observed  in the field of the MSP in infrared, to be part of a HT, while incorporating the effects due to the quadrupolar interactions between the binary orbits and the dominant contributions to the general relativistic periastron precession in the inner binary \citep{BLS02}.
Using results from the timing of the MSP, we demonstrate that the eccentricity of the binary pulsar should decrease at the present epoch and estimate the rate to be $\sim 10^{-13}{\rm s^{-1}} $. As this is clearly inconsistent with the measured $\dot e \sim 10^{-16}{\rm s^{-1}} $ for the binary (Freire, priv. commun.), we rule out Kozai resonance as the reason for the high orbital eccentricity for the binary MSP PSR J1903+0327.
 
\section*{Acknowledgments}

We thank Michael Kramer for the detailed referee report and Paulo Freire for sharing recent timing measurements of J1903+0327 prior to publication. AG is grateful to Gerhard Sch\"afer for enlightening discussions. AR thanks Daniel Fabrycky for discussions on spin period evolution. This work is supported in parts by grants from  the Deutsche Forschungsgemeinschaft (DFG) through SFB/TR7 ``Gravitationswellenastronomie'' and DLR (Deutsches Zentrum f\"ur Luft- und Raumfahrt) to AG. At TIFR this research is part of Eleventh Five Year Plan Projects numbered 11P-407 and 11P-409.

\label{lastpage}

\begin{thebibliography}{99}

\bibitem[Alpar $et~al.$ 1982]{Alpar_82} Alpar, M. A., Cheng, A. F., Ruderman, M. A. \&
Shaham, J., 1982, Nature, 300, 728.

\bibitem[Bagchi \& Ray 2009a]{bagray09a}Bagchi, M., Ray, A., 2009a, ApJ, 693, L91.

\bibitem[Bagchi \& Ray 2009b]{bagray09b}Bagchi, M., Ray, A., 2009b, ApJ, 701, 1161.

\bibitem[Blaes $et~al.$ 2002]{BLS02}Blaes, O., Lee, M. H., Socrates, A., 2002, ApJ, 578, 775.

\bibitem[Bergeron, Saumon, Wesemael 1995]{ber95} Bergeron, P., Saumon, D., Wesemael, F., 1995, ApJ, 443, 764. 

\bibitem[Cardelli, Clayton \& Mathis 1989]{car89} Cardelli, J. A., Clayton, G. C., Mathis, J. S., 1989, ApJ, 345, 245.

\bibitem[Champion $et~al.$ 2008]{djc08} Champion, D. J., Ransom, S. M., Lazarus, P. $et~al.$, 2008, Science, 320, 1309.

\bibitem[Damour \& Deruelle 1985]{DD86}Damour, T., Deruelle, N., 1986, Ann. Inst, Henri Poincare Phys. Theor, 44, 263.

\bibitem[Ford $et~al.$ 2000]{FKR00}Ford, E. B., Kozinsky, B.; Rasio, F. A., 2000, ApJ, 535, 385.

\bibitem[Freire, priv. commun.]{fre09} Freire, P., private communication.

\bibitem[Joshi \& Rasio 1997]{JR97} Joshi, K.~J., \& Rasio, F.~A.\ 1997, ApJ, 488, 901. 

\bibitem[Kozai 1962]{K62} Kozai, Y., 1962, AJ, 67, 591.

\bibitem[L\"ohmer, Kramer, Driebe $et~al.$ 2004]{loh04} L\"ohmer, O., Kramer, M., Driebe, T., $et~al.$, 2004, A \&A, 426, 631.

\bibitem[Mardling \& Aarseth 2001]{MA_01}Mardling, R. A., Aarseth, S. J., 2001, MNRAS, 321, 398

\bibitem[Mazeh \& Shaham 1979]{MS79}Mazeh, T.; Shaham, J., 1979, A \& A, 77, 145.

\bibitem[Phinney 1992]{Phinney_92} Phinney, E. S., 1992, Phil. Trans. R. Soc. Lond., 341, 39.

\bibitem[Taylor \& Weisberg 1989]{TW} Taylor, J. H., Weisberg, J. M., 1989, ApJ, 345, 434.

\bibitem[Wex 1998]{wex98}Wex, N., 1998, MNRAS, 298, 67.

\end{thebibliography}
\end{document}